\documentclass[sigconf]{acmart}

\usepackage{booktabs} 
\usepackage{acronym}
\usepackage{framed}
\usepackage{subfigure}
\usepackage{multirow}
\usepackage{tabularx}
\usepackage{algpseudocode}
\usepackage{float}
\floatstyle{ruled}
\newfloat{algorithm}{thp}{lop}
\floatname{algorithm}{Algorithm}

\setcopyright{none}

\acmDOI{}

\acmISBN{}

\acmConference[]{}{}{} 
\acmYear{2018}
\copyrightyear{2018}

\acmPrice{}

\begin{document}

\newcommand{\etal}{~\textit{et al.}}
\newcommand{\AUC}{\textit{AUC}}
\newcommand{\FMEAS}{\textit{F-measure}}
\newcommand{\GMEAS}{\textit{G-measure}}
\newcommand{\MCC}{\textit{MCC}}
\newcommand{\AUCEC}{\textit{AUCEC}}
\newcommand{\RELB}{$RelB_{20\%}$}
\newcommand{\NECM}{$NECM_{15}$}
\newcommand{\RANKSCORE}{\textit{rankscore}}
\newcommand{\RANKSCORES}{\textit{rankscores}}
\newcommand{\RECALL}{\textit{recall}}
\newcommand{\PRECISION}{\textit{precision}}
\newcommand{\ERROR}{\textit{error}}

\title{Benchmarking cross-project defect prediction approaches with costs metrics}

\author{Steffen Herbold}
\orcid{0000-0001-9765-2803}
\affiliation{%
  \institution{University of Goettingen, Insititute of Computer Science}
  \city{G\"{o}ttingen}
  \country{Germany}
}
\email{herbold@cs.uni-goettingen.de}


\renewcommand{\shortauthors}{S. Herbold}

\acrodef{ANOVA}{ANalysis Of VAriance}
\acrodef{AST}{Abstract Syntax Tree}
\acrodef{AUC}{Area Under the ROC Curve}
\acrodef{Ca}{Afferent Coupling}
\acrodef{CBO}{Coupling Between Objects}
\acrodef{CCA}{Canonical Correlation Analysis}
\acrodef{CD}{Critical Distance}
\acrodef{Ce}{Efferent Coupling}
\acrodef{CFS}{Correlation-based Feature Subset}
\acrodef{CLA}{Clustering and LAbeling}
\acrodef{CODEP}{COmbined DEfect Predictor}
\acrodef{CPDP}{Cross-Project Defect Prediction}
\acrodef{DBSCAN}{Density-Based Spatial Clustering}
\acrodef{DCV}{Dataset Characteristic Vector}
\acrodef{DTB}{Double Transfer Boosting}
\acrodef{fn}{false negative}
\acrodef{fp}{false positive}
\acrodef{GB}{GigaByte}
\acrodef{HL}{Hosmer-Lemeshow}
\acrodef{ITS}{Issue Tracking System}
\acrodef{JIT}{Just In Time}
\acrodef{LCOM}{Lack of COhession between Methods}
\acrodef{LOC}{Lines Of Code}
\acrodef{MDP}{Metrics Data Program}
\acrodef{MI}{Metric and Instances selection}
\acrodef{MODEP}{MultiObjective DEfect Predictor}
\acrodef{MPDP}{Mixed-Project Defect Prediction}
\acrodef{NN}{Nearest Neighbor}
\acrodef{PCA}{Principle Component Analysis}
\acrodef{RAM}{Random Access Memory}
\acrodef{RFC}{Response For a Class}
\acrodef{SCM}{SourceCode Management system}
\acrodef{SVM}{Support Vector Machine}
\acrodef{TCA}{Transfer Component Analysis}
\acrodef{tn}{true negative}
\acrodef{tp}{true positive}
\acrodef{RBF}{Radial Basis Function}
\acrodef{ROC}{Receiver Operating Characteristic}
\acrodef{UMR}{Unified Metric Representation}
\acrodef{VCB}{Value-Cognitive Boosting}
\acrodef{WPDP}{Within-Project Defect Prediction}

\begin{abstract}
Defect prediction can be a powerful tool to guide the use of quality assurance
resources. In recent years, many researchers focused on the problem of
\ac{CPDP}, i.e., the creation of prediction models based on training data from
other projects. However, only few of the published papers evaluate the cost
efficiency of predictions, i.e., if they save costs
if they are used to guide quality assurance efforts. Within this paper, we provide a
benchmark of 26 \ac{CPDP} approaches based on cost metrics.
Our benchmark shows that trivially assuming everything as
defective is on average better than \ac{CPDP} under cost considerations.
Moreover, we show that our ranking of approaches using cost metrics is uncorrelated to a ranking based on metrics that do not directly consider
costs. These findings show that we must put more effort into evaluating the
actual benefits of \ac{CPDP}, as the current state of the art of \ac{CPDP} can
actually be beaten by a trivial approach in cost-oriented evaluations.
\end{abstract}

%
%



\maketitle

\acresetall

\section{introduction}
\label{sec:introduction}

Software defect prediction has been under investigation in our community for many
years. The reason for this is the huge cost saving potential of accurately
predicting in which parts of software defects are located. This information
can be used to guide quality assurance efforts and ideally have fewer post
release defects with less effort. Since one major problem
of defect prediction approaches is the availability of training data,
researchers turned towards \ac{CPDP}, i.e., the prediction of defects based on
training data from other projects. 
A recent benchmark by Herbold\etal~\cite{Herbold2017b} compared the performance of 24 suggested approaches by researchers between 2008 and 2015.
However, the benchmark has one major limitation: it does not consider the impact
on costs directly. Instead, Herbold\etal~ used the machine learning
metrics \AUC, \FMEAS, \GMEAS, and \MCC. From a cost perspective, these metrics
only make sense, if all software entities that are predicted as defective 
get additional quality assurance attention. Even then, the cost of missing
defects would be assumed as equal to the cost of additional review effort. However, according to the
literature, post-release defects cost 15-50 times more than additional quality
assurance effort~\cite{Khoshgoftaar2004}. Rahman\etal~\cite{Rahman2012} already
found, that the results regarding \ac{CPDP} may be very different, if considered
from a cost-oriented perspective. Whether this finding by
Rahman\etal~ translates to the majority of the \ac{CPDP} literature is unclear,
because a cost-sensitive evaluation is missing. 

We close this gap with this paper. Our contribution is an adoption of the
benchmark from Herbold\etal~\cite{Herbold2017b} for cost metrics to determine a
cost-sensitive ranking of \ac{CPDP} approaches.
We compare 26 \ac{CPDP} approaches published between 2008 and 2016, as well as
three baselines.
Our results show that a trivial baseline approach that considers all code as
defective is almost never significantly outperformed by \ac{CPDP}. Only one
\ac{CPDP} proposed by Liu\etal~\cite{Liu2010} performs better for one out of
three cost metrics, but even this is only the case on one of the two data sets
we use. For the other data set, the performance of not
statistically significantly better. For the other two cost metrics, no \ac{CPDP}
approach performs statistically significantly better than the trivial baseline. 
Moreover, there is no correlation between ranking \ac{CPDP} approaches based on
cost metrics, and the ranking produced by Herbold\etal~\cite{Herbold2017b}. Hence, cost metrics should always
be considered in addition to other metrics, because otherwise we do not evaluate
if proposed models to what they should: reduce costs.

The remainder of this paper is structured as follows. We discuss related work in
Section~\ref{sec:relatedwork}. Then, we introduce our benchmark methodology
including the research questions, data used, performance metrics, and
statistical evaluation in Section~\ref{sec:methodology}. Afterwards, we present
our results in Section~\ref{sec:results}, followed the discussion in
Section~\ref{sec:discussion} and the threats to validity in
Section~\ref{sec:threats}.
Finally, we conclude the paper in Section~\ref{sec:conclusion}. 

\section{Related Work}
\label{sec:relatedwork}

\begin{table*}
\caption{Related work on \ac{CPDP} included in the benchmark.
Acronyms are defined following the authors and reference in the first column.}
\label{tbl:relwork}
\vspace{-8pt}
\begin{center}
\begin{tabular}{ p{0.16\linewidth}  p{0.8\linewidth} }
\hline
\textbf{Acronym} & \textbf{Short Description} \\
\hline
Khoshgoftaar08 & Khoshgoftaar\etal~\cite{Khoshgoftaar2008}
proposed majority voting of multiple classifiers trained for each product in the
training data.\\
Watanabe08 & Watanabe\etal~\cite{Watanabe2008} proposed standardization based
on mean values of the target product. \\
Turhan09 & Turhan\etal~\cite{Turhan2009} proposed a log-transformation and
nearest neighbor relevancy filtering. \\
Zimmermann09 & Zimmermann\etal~\cite{Zimmermann2009} proposed a decision
tree to select suitable training data.\\
CamargoCruz09 & Camargo Cruz and Ochimizu~\cite{CamargoCruz2009}
proposed a log-transformation and standardization based on the median of the
target product.\\
Liu10 & Liu\etal~\cite{Liu2010} proposed an S-expression tree created by
genetic program. \\
Menzies11 & Menzies\etal~\cite{Menzies2011,Menzies2013} proposed local models
for different regions of the training data through clustering.\\
Ma12 & Ma\etal~\cite{Ma2012} proposed data weighting using the concept of
gravitation. \\
Peters12 & Peters and Menzies~\cite{Peters2012} proposed an approach for data
privacy using a randomized transformation called MORPH. \\
Uchigaki12 & Uchigaki\etal~\cite{Uchigaki2012} proposed an ensemble of
univariate logistic regression models build for each attribute separately. \\
Canfora13 & Canfora\etal~\cite{Canfora2013,Canfora2015} proposed
a multi-objective genetic program to build a logistic regression model that
optimizes costs and the number of defects detected. \\
Peters13 & Peters\etal~\cite{Peters2013a} proposed relevancy filtering using
conditional probabilities combined with MORPH data privatization. \\
Herbold13 & Herbold~\cite{Herbold2013} proposed relevancy filtering using
distributional characteristics of products. \\
ZHe13 & Z. He\etal~\cite{He2013} proposed attribute selection and relevancy
filtering using separability between training and target products.  \\
Panichella14 & Panichella\etal~\cite{Panichella2014} proposed the CODEP
meta classifier over the results of multiple classification models. \\
Ryu14 & Ryu\etal~\cite{Ryu2014} proposed similarity based resampling and
boosting. \\
PHe15 & P. He\etal~\cite{He2015} proposed feature selection based on how
often the metrics are used for classification models build using the training
data. \\
Peters15 & Peters\etal~\cite{Peters2015} proposed LACE2 as an extension
with further privacy of CLIFF and MORPH. \\
Kawata15 & Kawata\etal~\cite{Kawata2015} proposed relevancy filtering
using DBSCAN clustering. \\
YZhang15 & Y. Zhang\etal~\cite{Zhang2015a} proposed the
ensemble classifiers average voting, maximum voting, boosting
and bagging. \\
Amasaki15 & Amasaki\etal~\cite{Amasaki2015} proposed feature selection and
relevancy filtering based on minimal metric distances between training and
target data. \\
Ryu15 & Ryu\etal~\cite{Ryu2015b} proposed relevancy filtering based on string
distances and LASER classification. \\
Nam15 & Nam and Kim~\cite{Nam2015b} proposed unsupervised defect
prediction based on the median of attributes. \\
Tantithamthavorn16 & Tantithamthavorn\etal, 2016~\cite{Tantithamthavorn2016}
proposed to use hyper parameter optimization for classifiers. \\
FZhang16 & F. Zhang\etal~\cite{Zhang2016} proposed unsupervised defect
prediction based on spectral clustering. \\
Hosseini16 & Hosseini\etal~\cite{Hosseini2016,Hosseini2017} proposed a genetic
program and nearest neighbor relevancy filtering to select training data. \\
\hline
\end{tabular}
\end{center}
\vspace{-8pt}
\end{table*}

We split our discussion of the related work into two parts. First, we discuss
the related work on defect prediction benchmarks. Second, we discuss the related
work on \ac{CPDP}.

\subsection{Defect prediction benchmarks}

Our benchmark on cost aspects is influenced by four defect prediction benchmarks
from the literature. Lessmann\etal~\cite{Lessmann2008} and
Ghotra\etal~\cite{Ghotra2015} evaluated the impact of different classifiers on
\ac{WPDP}. D'Ambros\etal~\cite{DAmbros2012} compared different kinds of metrics
and classification models for \ac{WPDP}. Herbold\etal~\cite{Herbold2017b}
compared \ac{CPDP} approaches on multiple data sets using multiple performance metrics. 

The benchmarks by Lessmann\etal, D'Ambros\etal, and Herbold\etal~ followed
Dem\v{s}ar's guidelines~\cite{Demsar2006} and make use of the Friedman
test~\cite{Friedman1940} with the post-hoc Nemenyi test~\cite{Nemenyi1963}.
Herbold\etal~ extended this concept using \RANKSCORES, that allow the
combination of results from multiple data sets and performance metrics into a single ranking.
Ghotra\etal~ use a different statistical procedure based on
ANOVA~\cite{Friedman1940} and the Scott-Knott test~\cite{Scott1974}.

Our benchmark design is similar to the design used by
Herbold\etal~\cite{Herbold2017b}. However, there are two major differences: 1)
we focus on different research questions, i.e., the performance using
cost-metrics, whereas Herbold\etal~ focused on machine learning metrics that do
not take costs into account; and 2) we extended the statistical evaluation with
a recently proposed effect size correction taking pattern from
ScottKnottESD proposed by Tantithamthavorn\etal~\cite{Tantithamthavorn2017}.

\subsection{Cross-project defect prediction}

The scope of our benchmark are approaches that predict software defects in a
target product using metric data collected from other projects. However, our
benchmark does not cover the complete body of \ac{CPDP} research. Specifically,
we do not address the following.
\begin{itemize}
  \item Mixed-project defect prediction, i.e., approaches that require labelled
  training data from the target product, e.g., \cite{Turhan2011, Turhan2013,
  Chen2015, Ryu2015a, Xia2016}.
  \item Heterogenuous defect prediction, i.e., approaches that work with
  different metric sets for the training and target products, e.g.,
  \cite{Nam2017, Jing2015}. 
  \item Just-in-time defect prediction, i.e., defect prediction for specific
  commits, e.g., \cite{Fukushima2014, Kamei2015}.
  \item Approaches that require project context factors, e.g., \cite{Zhang2015}.
\end{itemize}

Additionally, we excluded two works that use transfer component
analysis~\cite{Pan2011} by Nam\etal~\cite{Nam2013} and by
Jing\etal~\cite{Jing2017}. Transfer component analysis has
major scalability issues due to a very large eigenvalue problem that needs to
be solved. Herbold\etal~\cite{Herbold2017} already determined that they could only compute
results in less than one day for the smaller data sets for the approach by
Nam\etal~\cite{Nam2013}. We tried to resolve this problem by using a scientific
compute cluster, where we had access to nodes with up to 64 cores
and 256 GB of memory. We were still not able to
compute results for a large data set with 17681 instances\footnote{JURECZKO
as defined in Section~\ref{sec:data}} before hitting the execution time limit for
computational jobs of 48 hours.

This leaves us with 26 approaches that were published through 29 publications
listed in Table~\ref{tbl:relwork}. For each of these approaches, we define an
acronym, which we will use hereafter to refer to the approach and a short
description of the approach. The list in Table~\ref{tbl:relwork} is mostly consistent with
the benchmark from Herbold\etal~\cite{Herbold2017b} on \ac{CPDP}. Additionally,
we extended the work by Herbold\etal~ with three further replications of
approaches that were published in 2016, i.e., Tantithamthavorn16,
FZhang16, and Hosseini16.

\section{Benchmark Methodology}
\label{sec:methodology}

We now describe the methodology of our benchmark, including our research
questions, the data we used, the baselines and classifiers, the performance
metrics, and the statistical analysis. 

\subsection{Research Questions}

We address the following six research questions within this study. 

\begin{itemize}
  \item[RQ1:] Does it matter if we use defect counts or binary labels for cost
  metrics?
  \item[RQ2:] Which approach performs best if quality assurance is
  applied according to the prediction?
  \item[RQ3:] Which approach performs best if additional quality assurance
  can only be applied to a small portion of the code?
  \item[RQ4:] Which approach performs best independent of the prediction
  threshold?
  \item[RQ5:] Which approach performs best overall in terms of costs?
  \item[RQ6:] Is the overall ranking based on cost metrics different
  from the overall ranking based on the \AUC, \FMEAS, \GMEAS, and \MCC? 
\end{itemize}

With RQ1, we address the question if binary labels as defective/non-defective
are sufficient, or if defect counts are required to compare costs.
Binary labels carry less information and should lead to less accurate results.
For example, you save more costs if you prevent two defects, instead of one. 
The question is, does this really matter, i.e., do the values of performance
metrics change significantly? This question is especially interesting, as not
all publicly available defect prediction data sets provide defect counts. In
case the impact of defect counts is large, we may only use data sets
that provide this information for our benchmark.

With research questions RQ2-RQ4 we consider different quality assurance
scenarios. RQ2 explores the costs savings, if someone trusts
the defect prediction model completely, i.e., applies quality assurance measures
exactly according to the prediction of the model. RQ3 considers the case where the defect prediction
model is used to identify a small portion of the code for additional quality
assurance, i.e., a setting with a limited quality assurance budget. RQ2 and RQ3
have specific prediction thresholds, i.e., a certain amount of the predicted
defects are considered. With RQ4, we provide a threshold independent view on
costs, which is valuable for selecting an approach if the amount of effort to be
invested is unclear beforehand. To provide a general purpose ranking, we throw
all the considerations from RQ2-RQ4 together for RQ5 and evaluate which approach
performs best if all criteria are considered. Thus, RQ5 provides an evaluation
that accounts for different application and cost scenarios. With RQ6 we address
the question if the usually used machine learning metrics are sufficient to
estimate a cost-sensitive ranking, i.e., if they produce a similar or a
different ranking from directly using cost metrics. 

\subsection{Data}
\label{sec:data}

\begin{table}
\caption{Summary of used data sets.}
\label{tbl:datasets}
\vspace{-8pt}
\begin{tabular}{lllll}
\hline
\textbf{Name} & \textbf{\#Products} & \textbf{\#Instances} &
\textbf{\#Defective} \\
\hline
JURECZKO & 62 & 17681 & 6062  \\
AEEEM & 5 & 5371 & 893  \\
\hline
\end{tabular}
\vspace{-12pt}
\end{table}

We use data from two different defect prediction data sets from the literature
for our benchmark listed in Table~\ref{tbl:datasets}. These data sets are a
subset of the data sets that Herbold\etal~\cite{Herbold2017b} used in
their benchmark. The other three data sets that Herbold\etal~ used could not be
used for different reasons. The MDP and RELINK data were infeasible due to our
results regarding RQ1 (see Section~\ref{sec:results}). The NETGENE data does not
contain the size of artifacts, which is required for cost-sensitive
evaluations.\footnote{Accoding to the paper by Herzig\etal~\cite{Herzig2013},
complexity and size should be included in the data. There are archives called
complexity\_diff.tar.gz available for each product in the data. However, the
archives seem to contain the differences for complexity and size metrics for
each transaction identified by a revision hash. We did not find absolute values
for the size, which would be required for cost metrics.}

We can only give a brief summary of both data sets due to space
restrictions. Full lists of the software metrics, products contained, etc. can
be found in the literature at the cited references for each data set or
summarized in the benchmark by Herbold\etal~\cite{Herbold2017b}.

The first data set was donated by Jureczko and
Madeyski~\cite{Jureczko2010}\footnote{The data is publicly available online:
\url{http://openscience.us/repo/defect/ck/} (last checked: 2017-08-25).} and
consists of 48 product releases of 15 open source projects, 27 product
releases of six proprietary projects and 17 academic products that were
implemented by students. For each of these releases, 20 static product metrics
for Java classes, as well as the number of defects are part of the data. Taking
pattern from Herbold\etal~\cite{Herbold2017}, we use 62 of the products.
We do not use the 27 proprietary products to avoid threats to the
validity of our results due to mixing proprietary and open source software.
Moreover, three of the academic products contain less than five
defective instances, which is too few for reasonable analysis with machine
learning. In the following, we will refer to this data set as JUREZCKO.  

The second data set was published by
D'Ambros\etal~\cite{DAmbros2010}\footnote{The data is publicly available online:
\url{http://bug.inf.usi.ch/} (last checked: 2017-08-25)} and consists of five
software releases from different projects. For each of these releases, 71
software metrics for Java classes are available, that include static product
metrics and process metrics, like weighted churn and linearly decayed entropy, as well as the number
of defects. In the following, we refer to this data set as AEEEM, taking pattern
from Nam\etal~\cite{Nam2013}. 

\subsection{Baselines and Classifiers}

Because we adopt Herbold\etal's~\cite{Herbold2017b} benchmark methodology for
using cost metrics, our choices for baselines and classifiers are
nearly identical. We adopt three performance baselines from
Herbold\etal~\cite{Herbold2017b} that define na\"{i}ve approaches for
classification models: ALL that takes all available training data as is, RANDOM
which randomly classifies instances as defective with a probability of 0.5,
and FIX which classifies all instances as defective. We do not adopt the
baseline CV for 10x10 cross-validation, because cross-validation is not
implementable in practice and is known to overestimate the performance of
\ac{WPDP}~\cite{Tan2015}. This may skew rankings, as approaches may be
outperformed by something that overestimates performance. Moreover,
cross-validation is an estimator for \ac{WPDP} performance and, therefore, out
of scope of our benchmark for \ac{CPDP} models.

We use a C4.5 decision tree (DT)~\cite{Quinlan1993}, logistic
regression (LR)~\cite{Cox1958}, na\"{i}ve bayes (NB)~\cite{Russell2003}, random
forest (RF)~\cite{Breiman2001}, RBF
network~(NET)~\cite{Caruana2006,Broomhead1988}, and a support vector machine
with radial basis function kernel (SVM)~\cite{Gestel2004} for all approaches that did not propose a classifier,
but rather a treatment of the training data or something similar. These are the
sixteen approaches Koshgoftaar08, Watanabe08, Turhan09, Zimmermann09,
CamargoCruz09, Ma12, Peters12, Peters13, Herbold13, ZHe13, PHe15, Peters15,
Kawata15, Amasaki15, Ryu15, and Nam15, as well as the baseline ALL.

For Menzies11, we use the WHICH classifier, which was used in the
original publication together with the six classifiers from above (DR, LR, NB,
RF, NET, and SVM), i.e., a total of seven classifiers. For Tantithamthavorn16 we
apply the proposed hyper parameter optimization to NB, RF, and SVM with the
same parameters as suggested by
Tantithamthavorn\etal~\cite{Tantithamthavorn2016}. For DT, NET, and LR no hyper
parameters to optimize are contained in the caret R package~\cite{Kuhn2016}
suggested by Tantithamthavorn\etal Additionally, we train a C5.0
decision tree~\cite{Kuhn2015} with hyper parameter optimization, because it
performs best in the evaluation by Tantithamthavorn\etal. We refer to these as
optimized classifiers as NBCARET, RFCARET, SVMCARET and C50CARET. 

The remaining eight of the approaches directly propose a classification
scheme, which we use: 
\begin{itemize}
  \item genetic program (GP) for Liu10;
  \item logistic ensemble (LE) for Uchigaki12;
  \item MODEP for Canfora13;
  \item CODEP with Logistic Regression (CODEP-LR) and CODEP with a Bayesian
  Network (CODEP-BN) for Panichella14;
  \item the value-cognitive boosted SVM (VCBSVM) for Ryu14; 
  \item average voting (AVGVOTE), maximum voting (MAXVOTE), bagging with a C4.5
  Decision Tree (BAG-DT), bagging with Na\"{i}ve Bayes (BAG-NB), boosting with a
  C4.5 Decision Tree (BOOST-DT), boosting with Na\"{i}ve Bayes (BOOST-NB) for
  YZhang15;
  \item spectral clustering (SC) for FZhang16; and
  \item search-based selection (SBS) for Hosseini16.
\end{itemize}

The MODEP classifier by Canfora13 requires either a constraint with a desired
\RECALL, or a desired cost objective. Herbold\etal~\cite{Herbold2017b} decided
to use a \RECALL~ of 0.7 as the constraint. In our benchmark, we sample
different values for \RECALL~ and use the values 0.1 to 1.0 in steps of 0.1. We denote the
different \RECALL~ constraints after the classifier name using the percentage,
e.g., MODEP10 for the constraint \RECALL=0.1. 

To deal with randomization, we repeat all approaches that contain random
components 10 times and then use the mean value of these repetitions for
comparison with the other approaches. Following
Herbold\etal~\cite{Herbold2017b}, these are Liu10, Canfora10, Menzies11,
Peters12, Peters13, ZHe13, Peters15, as well as the baseline RANDOM.
Moreover, two of the three approaches that we added to the benchmark 
contain random components, i.e., Hosseini16 because of the random test data splits and the genetic program and
Tantithamthavorn16 because of the cross-validation for the hyper parameter
optimization. 


\subsection{Performance Metrics}

To properly evaluate our research questions, we require metrics that
measure the cost for different settings. In order to not re-invent the wheel,
we scanned the literature and found fitting metrics for all our research
questions. 

For RQ2, we need a metric that can be used to evaluate the costs if one follows
the classification achieved with a defect prediction model, i.e., to apply
quality assurance ot everything that is predicted as defective and nothing else. We use the metric $NECM_{C_{ratio}}$, which is
defined as
\begin{equation}
NECM_{C_{ratio}} = \frac{fp + C_{ratio}\cdot fn}{tp+fp+tn+fn}
\end{equation}
This metric was, e.g., used by Liu\etal~\cite{Liu2010} and
Khoshgoftaar\etal~\cite{Khoshgoftaar2008} and measures the costs resulting
from overhead in quality assurance effort through false positive predictions
versus the costs due to missed defects through false negative predictions.
$C_{ratio}$ is used to define the difference in cost for false positive and false negative
predictions. Khoshgoftaar\etal~ and Liu\etal~ both use 15, 20, and 25 as values
for $C_{ratio}$. The cost of missing defects may be 15-50 times higher
than that of additional quality assurance measures
according to the literature~\cite{Khoshgoftaar2004}.
For our benchmark, we use $C_{ratio}=15$, i.e., the most conservative cost
scenario, where reviews are relatively expensive in comparison to the saved costs of finding a defect through the prediction
model. 

To evaluate RQ3, i.e, the costs when only a small part of the
code shall be reviewed, we use the metric \RELB, defined as the percentage of
defects found when inspecting 20\% of the code. Thus, the defect prediction
model is used to rank all code entities. Then, the entities are considered
starting with the highest ranked entity until 20\% of the code are covered. 
This metric is an adoption of
the metric $\textit{NofB}_{20\%}$ used by Y. Zhang\etal~\cite{Zhang2015a}. The
only difference is that Y. Zhang\etal~ consider absolute numbers, whereas we
consider the percentage. This difference is required due to the diversity in the
size of software products. If we do not remove the strong impact of the project
size from the values for this metric and use absolutes instead of ratios, our
statistical analysis would be strongly influenced by the size of the products
and not measure the actual defect detection capability.

The metrics \NECM~ and \RELB~ evaluate the \ac{CPDP} models in such a way that a
fixed amount of code is considered for additional quality assurance. 
For \NECM, we follow the classification, i.e., we consider the scenario that is
most likely according to the predictions of the \ac{CPDP} model. For \RELB, we
stop at 20\% of the code, regardless of the scores of the classification model.
This is related to the notion of thresholds for classification in the machine
learning world: a learned prediction model has a scoring function as output and
everything above a certain threshold is then classified as defective. While
there are good reasons to use the stratgies the thresholds are picked by \NECM~
and \RELB~, these thresholds are still magic numbers. 
To evaluate RQ4, we use the threshold independent metric \AUCEC, which is defined
as the area under the curve of review effort versus number of defects
found~\cite{Rahman2012}. This is a threshold independent metric that analyzes
the performance of a defect likelihood ranking produced by a classifier for all
possible thresholds. Thus, the value of \AUCEC~ is not a measure for a single
prediction model with a fixed threshold like \NECM~ and \RELB~, but instead for
a family of prediction models with all possible threshold values. Regarding
costs, this means that \AUCEC~ evaluates the costs for all possible amounts of
code that to which additional quality assurance is applied, starting from
applying no quality assurance at all and stopping with applying quality
assurance to the complete product.

To evaluate RQ1 and RQ5, we use the metrics \NECM, \RELB, and \AUCEC~
together.
For RQ6, we compare the findings of RQ5 to the results if we use the benchmark criteria from Herbold\etal~\cite{Herbold2017}. Thus we
determine how different the ranking from RQ5 is from a ranking using the
metrics \AUC, \FMEAS, \GMEAS, and \MCC. The definition and reasons for
selecting these metrics can be found in the benchmark by Herbold\etal~\cite{Herbold2017}. 

We use actual defect counts for the number of true positives
and false negatives to compute the above metrics. 
Thus, if we have an instance with two defects, it carries twice the weight for the above performance
metrics. The only exception to this is RQ1, where we compare using binary labels
to using defect counts. With binary labels, we do not care about the number of
defects in a class and just label it as defective or non-defective, meaning that classes
with five defects have the same weight as classes with a one defect.

In case of ties, i.e., two instances with the same score according to a
prediction model, we use the size of the instance as tie breaker and say that
the smaller instances get additional quality assurance first. This
tie-breaking strategy was proposed by Rahman\etal~\cite{Rahman2012}.

\subsection{Statistical Analysis}

We use a threshold of $\alpha=0.995$ for all statistical tests instead of the
usually used $\alpha=0.95$. Hence, we require that $\textit{p-value} < 0.005$
instead of the usual $\textit{p-value} < 0.05$ to reject a null hypothesis.
We decided to follow the recent suggestion made by a broad array of
researchers from different domains to raise the threshold for
significance~\cite{Benjamin2017}. The reason for this higher threshold is to
reduce the likelihood that approaches are false positively detected as
significantly different, even though they are not.

For RQ1, we compare the mean performance of each approach achieved using
binary labels for calculating metric values with numerical values for calculating the metrics. We
say that there is a difference if the mean value is statistically significantly
different and the effect size is non-negligible. For the testing of statistical
significance, we use the non-parametric Mann-Whitney-U test~\cite{Mann1947}.
In case the difference is statistically significant, we measure the effect size
using Cohen's $d$~\cite{Cohen1988}. According to Cohen, the effect size is
negligible for $d < 0.2$, small for $0.2 \leq d < 0.5$, medium for $0.5 \leq d < 0.8$, and
large for $d \geq 0.8$. We used Levene's test~\cite{Brown1974} to test if the
homoscedasticity assumption of Cohen's $d$ is fulfilled. In case we find that
binary labels lead to statistically significantly different results with a non-neglible effect
size, data sets with binary labels instead of defect counts should not be used
for cost-sensitive evaluations.

For the research questions RQ2-RQ6, we require a ranking of multiple approaches
and, thus, a more complex statistical testing procedure. Our first step was to
determine if we can use ANOVA~\cite{Friedman1940} in combination with a Scott-Knott
test~\cite{Scott1974}, which is a popular choice in recent defect prediction
literature that compares multiple approaches to each other, e.g.,
\cite{Ghotra2015, Herbold2016, Tantithamthavorn2016, Tantithamthavorn2017}.
The advantage of ANOVA and Scott-Knott is a clear and non-overlapping ranking of
results. However, ANOVA has the heavy assumptions that all populations follow a normal distribution and are homoscedastic. We
used the Shapiro-Wilk test~\cite{Shapiro1965} to test if the performance values are
normally distributed and used Levene's test~\cite{Brown1974} to test for
homoscedasticity. Unfortunately, both conditions are frequently broken by the
data. 

Therefore, we decided to use the less powerful but non-parametric
Friedmann test~\cite{Friedman1940} with the post-hoc Nemenyi
test~\cite{Nemenyi1963} instead. The Nemenyi test compares the
distances between the mean ranks of multiple pair-wise comparisons between all
approaches on all products of a data set. The main drawback of this test is that the
ranks between approaches may be overlapping. To deal with this issue, we follow
the strategy suggested by Herbold\etal~\cite{Herbold2017d} to create
non-overlapping groups of statistically significantly different results.
Herbold\etal~suggest to start with the best ranked approach and always create a
new group, if the difference in ranking between two subsequently ranked
approaches is greater then the critical distance. At the cost of discriminatory
power of the test, this ensures that the resulting groups are non-overlapping
and statistically significantly different.

Moreover, we took pattern from Tantithamthavorn\etal's modification of the
Scott-Knott test~\cite{Tantithamthavorn2017} and adopted the proposed effect
size correction. This means that we use Cohen's $d$ to measure the effect size
between two subsequently ranked groups and merge them if the effect size is
negligible, i.e., $d < 0.2$.  

The final step of the statistical analysis is the generation of the \RANKSCORE~
from the ranking. The \RANKSCORE~ was introduced by
Herbold\etal~\cite{Herbold2017b} to deal with the problem of different group
sizes that occur when ranks of groups of approaches are created. Not every group
will have the same number of approaches. This means that the number of the
group-ranking becomes a bad estimator for the performance of the group. 
Basically, it is a difference if you are in the second-ranked group and there
is one approach in the first-ranked group or there are ten approaches in the
first-ranked group. The \RANKSCORE~ takes care of this problem by transforming
the ranks into a normalized representation based on the percentage of approaches
that are on higher ranks, i.e., 
\begin{equation*}
\textit{rankscore} = 1-\frac{\#\{\text{approaches ranked
higher}\}}{\#\{\text{approaches}\}-1}.
\end{equation*}
For example, a $\RANKSCORE=1$ is perfect meaning that no approach is
ranked higher, a $\RANKSCORE=0.7$ would mean that 30\% of approaches are
ranked higher.

We apply this statistical evaluation procedure to each combination of data set
and performance measure. For research questions RQ2-RQ4, we then evaluate the mean
\RANKSCORE~ on both data sets for the metrics \NECM, \RELB, and \AUCEC,
respectively.
For RQ5, we evaluate the mean \RANKSCORE~ on both data sets and for all three
cost metrics. For RQ6, we evaluate the mean \RANKSCORE~ on both data sets for
the machine learning metrics \AUC, \FMEAS, \GMEAS, and \MCC. To evaluate the
relationship between the rankings produced with cost metrics and machine learning
metrics for RQ6, we evaluate the correlation between both using Kendall's
$\tau$~\cite{Kendall1948}. Kendall's $\tau$ is a non-parametric correlation
measure between ranks of results, i.e., the ordering of results produced, which
is exactly what we are interested in.


\section{Results}
\label{sec:results}

We now present the results of our benchmark. A replication kit that provides the
complete source code and data required for the replication of our results, a
tutorial on how to use the replication kit, as well as additional visualizations
including plots for that list all classifiers and not only the best classifiers,
and critical distance diagrams for the Nemenyi tests are available
online.\footnote{Reference to the replication kit removed due to double blind review. Will be uploaded to
Zenodo in case of acceptance and evaluated through the artifact evaluation. We
provide the visualiations and the statistical analysis code as supplemental
material to this submission.}

\begin{table}[t]
\centering
\caption{Mean performance with binary labels and defect counts, as well as the \textit{p-value} of the Mann-Whitney-U test.}
\label{tbl:numericimpact}
\vspace{-8pt}
\begin{tabular}{ll>{\centering\arraybackslash}p{1.2cm}>{\centering\arraybackslash}p{1.2cm}>{\centering\arraybackslash}p{1.2cm}}
  \hline
 \textbf{Dataset} & \textbf{Metric} & \textbf{Binary Label} & \textbf{Defect Counts} & \textit{p-value} ($d$) \\ 
  \hline
\multirow{3}{*}{AEEEM} & \NECM & 1.92 & 2.25 & 0.012 \\ 
  & \RELB & 0.28 & 0.29 & 0.288 \\ 
  & \AUCEC & 0.56 & 0.57 & 0.236 \\ 
  \hline
\multirow{3}{*}{JURECZKO} & \NECM & 3.44 & 4.10 & $<$0.001 (0.52) \\ 
  & \RELB & 0.23 & 0.21 & 0.631 \\ 
  & \AUCEC & 0.53 & 0.54 & 0.439 \\ 
   \hline
\end{tabular}
\vspace{-12pt}
\end{table}

\begin{figure*}
\caption{Mean \RANKSCORE~ over all data sets. The black diamonds depict the mean
\RANKSCORE, the gray points in the background are the \RANKSCORES~ over which
the mean is taken. We list only the result achieved with the best classifier for
each approach.}
\label{fig:rq2torq5}
\centering
\subfigure[Results with \NECM]{
\includegraphics[width=\columnwidth]{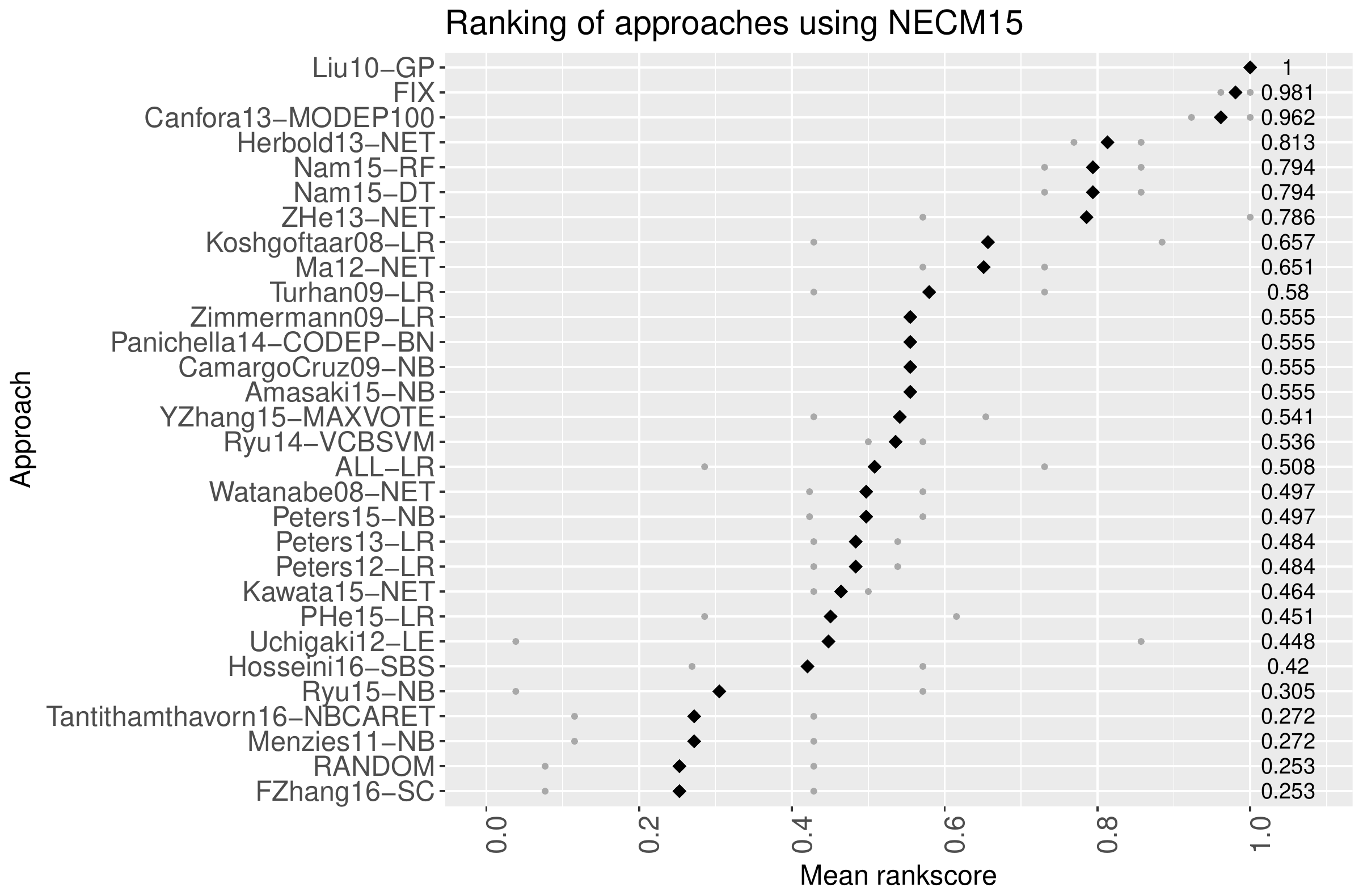}
\label{fig:rq2}
}
\subfigure[Results with \RELB]{
\includegraphics[width=\columnwidth]{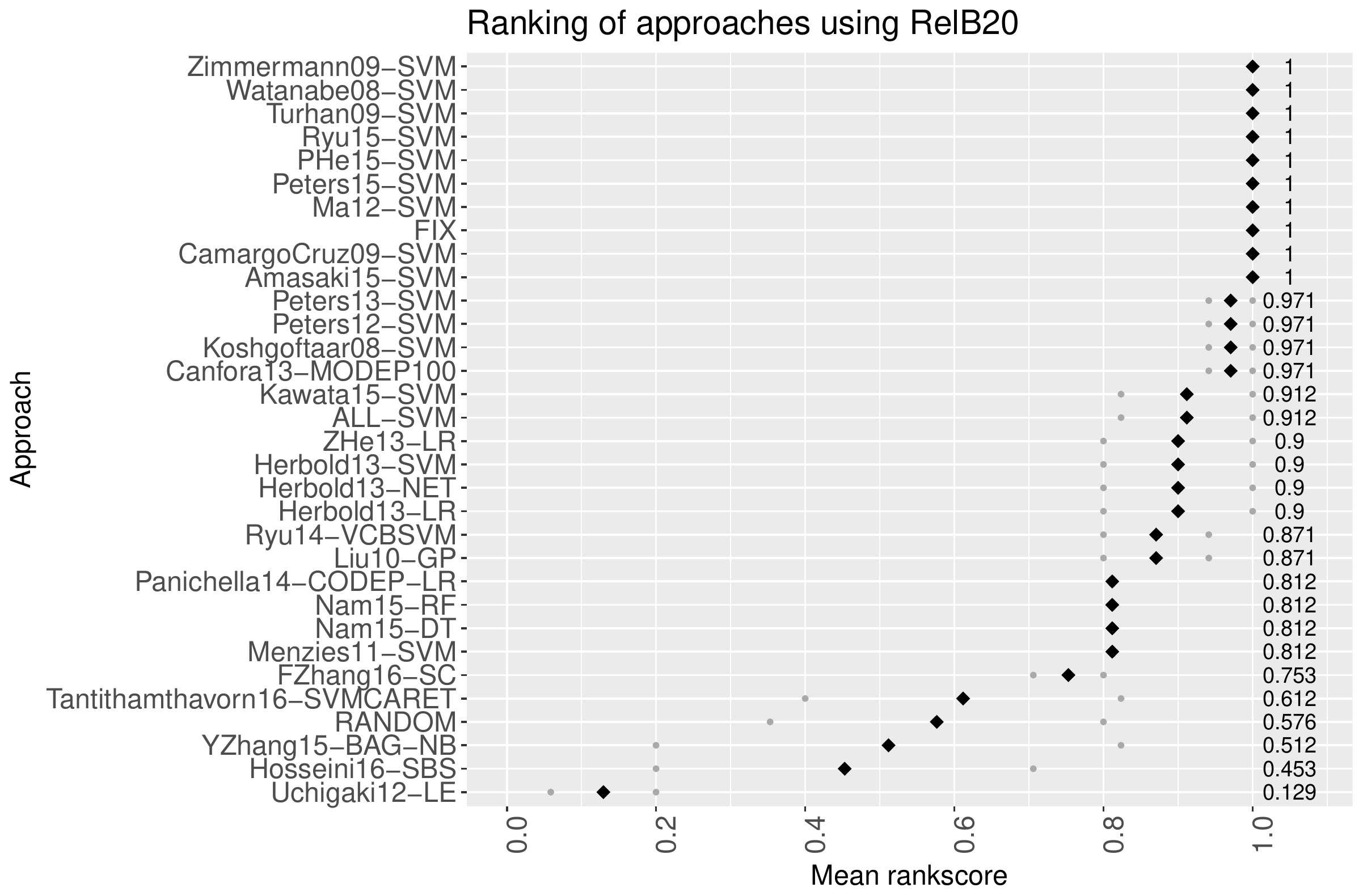}
\label{fig:rq3}
}
\subfigure[Results with \AUCEC]{
\includegraphics[width=\columnwidth]{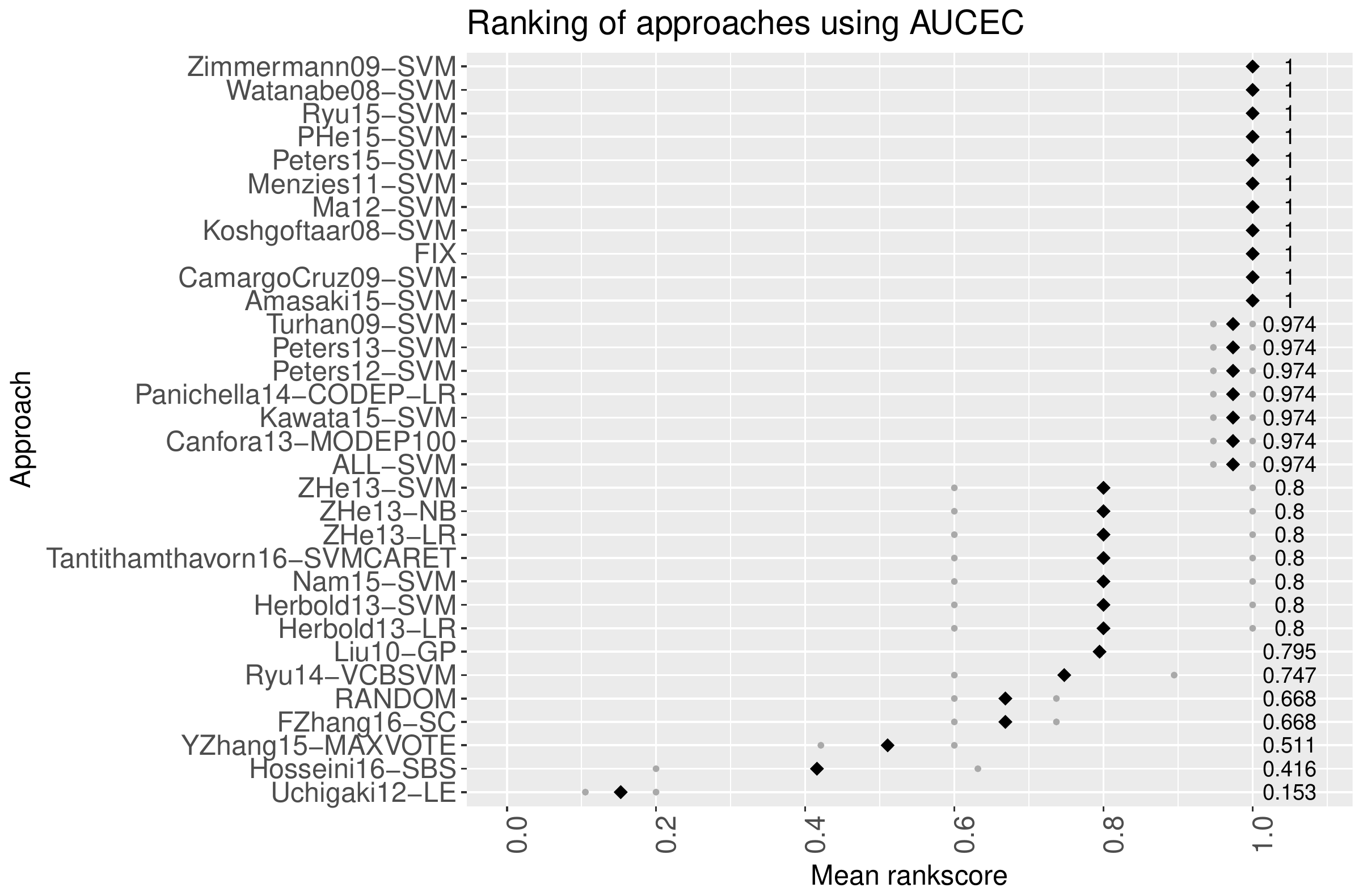}
\label{fig:rq4}
}
\subfigure[Results with \NECM, \RELB, and \AUCEC]{
\includegraphics[width=\columnwidth]{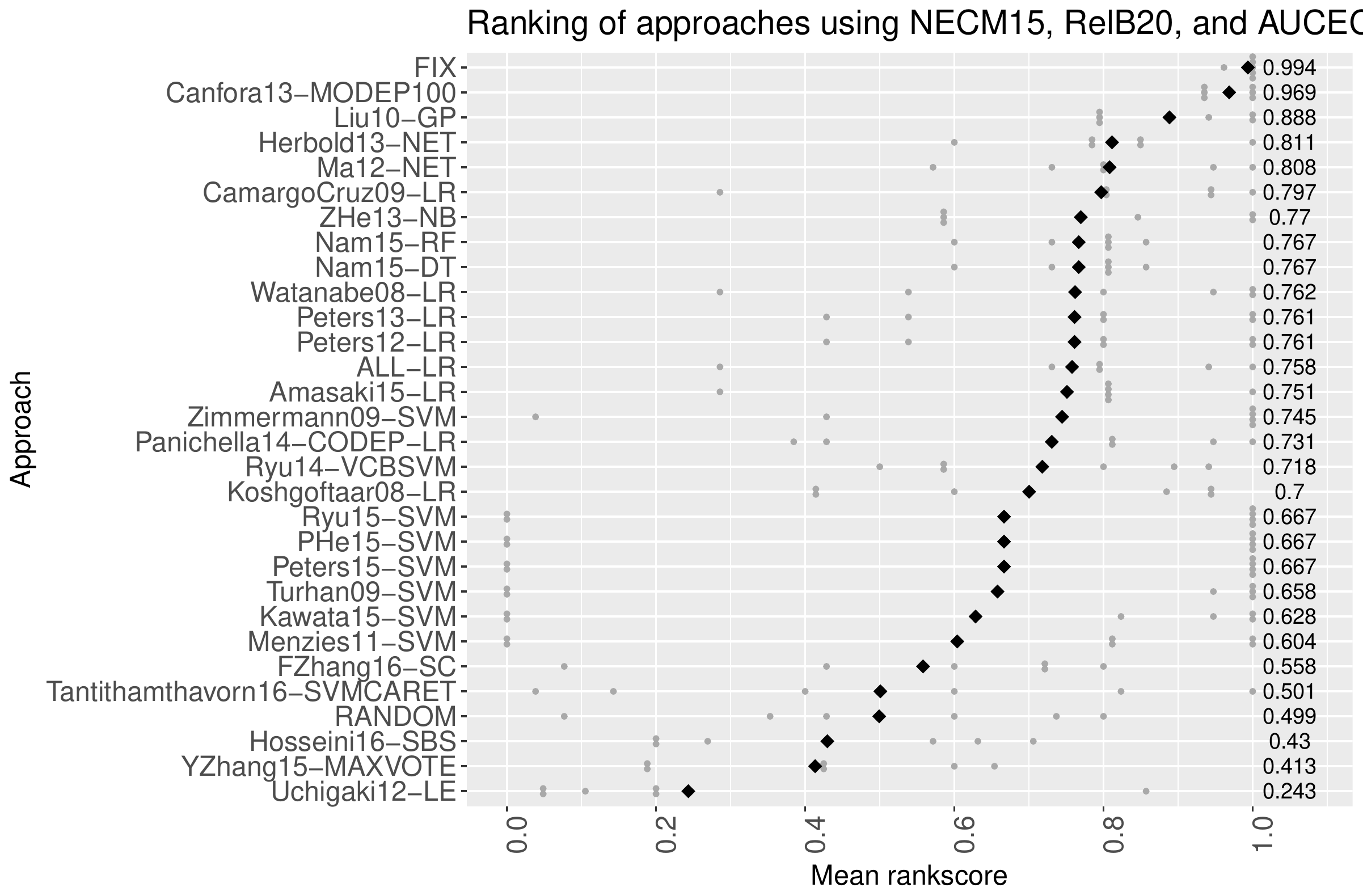}
\label{fig:rq5}
}
\vspace{-12pt}
\end{figure*}

\begin{figure}
\caption{Mean \RANKSCORE~ over all data sets for the metrics \AUC, \FMEAS,
\GMEAS, and \MCC. The black diamonds depict the mean \RANKSCORE, the gray points
in the background the \RANKSCORES~ over which the mean is taken. We list only
the results for the classifiers that are listed in Figure~\ref{fig:rq2torq5}(d), i.e., those
performing best according to the metrics \NECM, \RELB, and \AUCEC.}
\includegraphics[width=\columnwidth]{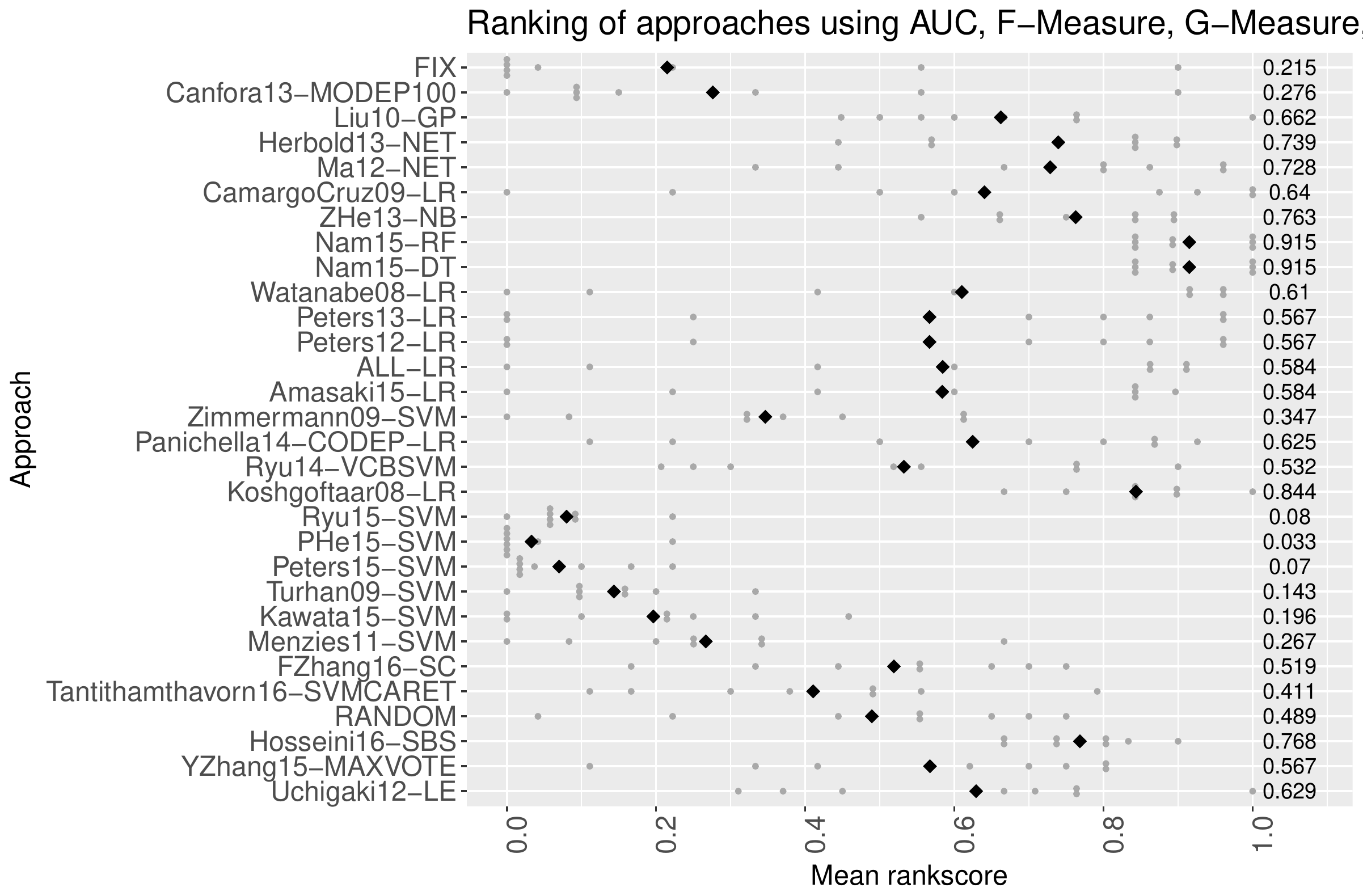}
\label{fig:rq6}
\vspace{-18pt}
\end{figure}

\subsection*{RQ1: Does it matter if we use defect counts or binary labels for
cost metrics?}

Table~\ref{tbl:numericimpact} shows the mean values for all cost metrics on
the two data sets where defect counts are available. We also report the
\textit{p-values} determined by the Mann-Whitney-U test, and in case of
significance, i.e., if $\textit{p-value}<0.005$, we also report the value for
Cohen's $d$ in parenthesis. We observe statistically significant differences for
\NECM~ on the JURECZKO data. Levene's test showed that the
results are homoscedastic. The effect size is $d=0.52$, i.e., medium.

\begin{framed}
\noindent\textbf{Answer RQ1:} For the metric \NECM, defect counts yield
significantly different results with a medium effect size in comparison to using
binary labels. Therefore, only data with defect counts should be used for
evaluations using \NECM. Consequently, we conclude that only
data with defect counts should be used for benchmarking with cost metrics and may not use
the MDP data and RELINK data used by Herbold\etal~\cite{Herbold2017b} for our
benchmark. 
\end{framed}

\subsection*{RQ2: Which approach performs best if quality assurance is
  applied according to the prediction?}
  
\begin{table}[t]
\caption{Mean results for \NECM~ with \RANKSCORE~ in parenthesis.}
\label{tbl:results-necm}
\vspace{-8pt}
\centering
\begin{tabular}{rll}\hline
\textbf{Approach} & \textbf{JURECZKO} & \textbf{AEEEM} \\ 
  \hline
ALL-LR & 4.7 (0.29) & 1.31 (0.73) \\ 
  Amasaki15-NB & 2.9 (0.57) & 1.56 (0.54) \\ 
  CamargoCruz09-NB & 2.92 (0.57) & 1.49 (0.54) \\ 
  Canfora13-MODEP100 & \textbf{0.55 (1)} & 0.76 (0.92) \\ 
  FZhang16-SC & 3.65 (0.43) & 2.35 (0.08) \\ 
  Herbold13-NET & 1.44 (0.86) & 0.91 (0.77) \\ 
  Hosseini16-SBS & 2.72 (0.57) & 1.63 (0.27) \\ 
  Kawata15-NET & 4.18 (0.43) & 1.61 (0.5) \\ 
  Koshgoftaar08-LR & 3.99 (0.43) & 1.13 (0.88) \\ 
  Liu10-GP & \textbf{1.3 (1)} & \textbf{0.82 (1)} \\ 
  Ma12-NET & 3.16 (0.57) & 1.27 (0.73) \\ 
  Menzies11-NB & 4.48 (0.43) & 2.07 (0.12) \\ 
  Nam15-DT & 1.94 (0.86) & 1.11 (0.73) \\ 
  Nam15-RF & 1.94 (0.86) & 1.11 (0.73) \\ 
  Panichella14-CODEP-BN & 3.66 (0.57) & 1.45 (0.54) \\ 
  Peters12-LR & 4.04 (0.43) & 1.41 (0.54) \\ 
  Peters13-LR & 4.04 (0.43) & 1.41 (0.54) \\ 
  Peters15-NB & 2.88 (0.57) & 1.73 (0.42) \\ 
  PHe15-LR & 5.36 (0.29) & 1.21 (0.62) \\ 
  RANDOM & 3.84 (0.43) & 2.48 (0.08) \\ 
  Ryu14-VCBSVM & 2.34 (0.57) & 1.71 (0.5) \\ 
  Ryu15-NB & 3.11 (0.57) & 3.27 (0.04) \\ 
  Tantitham.16-NBCARET & 3.82 (0.43) & 1.97 (0.12) \\ 
  FIX & \textbf{0.53 (1)} & 0.72 (0.96) \\ 
  Turhan09-LR & 3.88 (0.43) & 1.19 (0.73) \\ 
  Uchigaki12-LE & 1.56 (0.86) & 3 (0.04) \\ 
  Watanabe08-NET & 2.94 (0.57) & 1.57 (0.42) \\ 
  YZhang15-MAXVOTE & 3.49 (0.43) & 1.34 (0.65) \\ 
  ZHe13-NET & 2.71 (0.57) & \textbf{0.87 (1)} \\ 
  Zimmermann09-LR & 3.14 (0.57) & 1.57 (0.54) \\ 
   \hline
\end{tabular}
\vspace{-12pt}
\end{table}

Figure~\ref{fig:rq2torq5}(a) shows the best approaches ranked using \NECM~ by
their mean \RANKSCORE~ over both data sets. The mean value of \NECM\linebreak
and the \RANKSCORE~ for each of these best approaches are listed in Table~\ref{tbl:results-necm}. 
The best ranking approach is Liu10-GP with a perfect mean
\RANKSCORE~ of 1, i.e, for both data sets no
approach is significantly better. The trivial baseline FIX, i.e., predicting all
code as defective, is a close second with a mean \RANKSCORE~ of 0.981.
Thus, only one approach beats the trivial baseline FIX for this performance
metric. Another approach, Canfora-MODEP100 is very close to FIX with a
\RANKSCORE~ of 0.962.
If we look at the actual values of \NECM~ and not the \RANKSCORE, FIX actually
has better mean values than Liu10-GP for all three data sets, even though the
ranking is worse on the AEEEM data.\footnote{There are multiple such anomylies
in our results. We checked them and they are all due to the effect we describe
below and not due to problems with the statistical analyis.} This anomaly is
possible due to the nature of the Nemenyi test. The Nemenyi test is based on the
ranking in pairwise comparisons of all approaches, not on the mean value.
Table~\ref{tbl:nemenyi} shows \NECM~ value and ranking of both Liu10-GP and FIX
on each product in the AEEEM data. The equinox and the lucene
products are most interesting. On equinox, FIX has an advantage of 0.4 over
Liu10-GP with respect to the metric \NECM. However,
the difference in ranks to Liu10-GP is only three. On lucene, Liu10-GP has a only an
advantage of 0.12 over FIX with respect to the metric \NECM, but the
difference in ranks to FIX is 27. Thus, while this anomaly seems counter
intuitive, from a ranking perspective and also for the statistical test, it is
correct. Such effects are the reason why checking for assumptions and choosing
appropriate statistical tests is important, as these effects are due to the
heteroscedacity of the data.
This also shows that pure comparisons of characteristics like mean or median
values are not sufficient for the ranking of multiple approaches.

\begin{table}
\caption{Detailed results for \NECM on the AEEEM data for Liu10-GP and FIX.}
\label{tbl:nemenyi}
\vspace{-8pt}
\begin{tabular}{rllll}
\hline
 & \multicolumn{2}{c}{\textbf{Liu10-GP}} & \multicolumn{2}{c}{\textbf{FIX}} \\
\textbf{Product} & \NECM & \textbf{Rank} & \NECM & \textbf{Rank} \\
\hline
eclipse & 0.70 & 11 & 0.68 & 9\\ 
equinox & 0.84 & 5 & 0.44 & 2 \\
lucene & 0.75 & 21 & 0.87 & 38 \\
mylyn & 1.03 & 6 & 0.83 & 2 \\
pde & 0.78 & 1 & 0.79 & 5\\
\hline \hline
\textit{mean} & 0.82 & 8.8 & 0.72 & 11.2 \\ 
\hline
\end{tabular}
\end{table}

\begin{framed}
\noindent\textbf{Answer RQ2:} Liu10-GP yields the best cost performance assuming
missed defects are 15 times more expensive than additional quality assurance
costs through false positive predictions. The other 25 approaches perform worse
in terms of expected costs than trivially assuming that all code is defective,
although at least Canfora13-MODEP is very close to this trivial baseline.
\end{framed}

\subsection*{RQ3: Which approach performs best if additional quality assurance
  can only be applied to a small portion of the code?}
  
\begin{table}[t]
\caption{Mean results for \RELB~ with \RANKSCORE~ in parenthesis.}
\label{tbl:results-relb}
\vspace{-8pt}
\centering
\begin{tabular}{rll}\hline
\textbf{Approach} & \textbf{JURECZKO} & \textbf{AEEEM} \\ 
  \hline
ALL-SVM & \textbf{0.33 (1)} & 0.3 (0.82) \\ 
  Amasaki15-SVM & \textbf{0.34 (1)} & \textbf{0.35 (1)} \\ 
  CamargoCruz09-SVM & \textbf{0.34 (1)} & \textbf{0.35 (1)} \\ 
  Canfora13-MODEP100 & \textbf{0.33 (1)} & 0.32 (0.94) \\ 
  FZhang16-SC & 0.26 (0.8) & 0.28 (0.71) \\ 
  Herbold13-LR & 0.25 (0.8) & \textbf{0.36 (1)} \\ 
  Herbold13-NET & 0.26 (0.8) & \textbf{0.32 (1)} \\ 
  Herbold13-SVM & 0.24 (0.8) & \textbf{0.33 (1)} \\ 
  Hosseini16-SBS & 0.12 (0.2) & 0.29 (0.71) \\ 
  Kawata15-SVM & \textbf{0.34 (1)} & 0.3 (0.82) \\ 
  Koshgoftaar08-SVM & \textbf{0.34 (1)} & 0.32 (0.94) \\ 
  Liu10-GP & 0.27 (0.8) & 0.31 (0.94) \\ 
  Ma12-SVM & \textbf{0.33 (1)} & \textbf{0.32 (1)} \\ 
  Menzies11-SVM & 0.3 (0.8) & 0.31 (0.82) \\ 
  Nam15-DT & 0.24 (0.8) & 0.29 (0.82) \\ 
  Nam15-RF & 0.24 (0.8) & 0.29 (0.82) \\ 
  Panichella14-CODEP-LR & 0.3 (0.8) & 0.28 (0.82) \\ 
  Peters12-SVM & \textbf{0.33 (1)} & 0.32 (0.94) \\ 
  Peters13-SVM & \textbf{0.33 (1)} & 0.32 (0.94) \\ 
  Peters15-SVM & \textbf{0.34 (1)} & \textbf{0.33 (1)} \\ 
  PHe15-SVM & \textbf{0.34 (1)} & \textbf{0.32 (1)} \\ 
  RANDOM & 0.26 (0.8) & 0.25 (0.35) \\ 
  Ryu14-VCBSVM & 0.24 (0.8) & 0.31 (0.94) \\ 
  Ryu15-SVM & \textbf{0.33 (1)} & \textbf{0.32 (1)} \\ 
  Tantitham.16-SVMCARET & 0.2 (0.4) & 0.29 (0.82) \\ 
  FIX & \textbf{0.34 (1)} & \textbf{0.32 (1)} \\ 
  Turhan09-SVM & \textbf{0.32 (1)} & \textbf{0.33 (1)} \\ 
  Uchigaki12-LE & 0.11 (0.2) & 0.2 (0.06) \\ 
  Watanabe08-SVM & \textbf{0.32 (1)} & \textbf{0.33 (1)} \\ 
  YZhang15-BAG-NB & 0.11 (0.2) & 0.28 (0.82) \\ 
  ZHe13-LR & 0.23 (0.8) & \textbf{0.33 (1)} \\ 
  Zimmermann09-SVM & \textbf{0.34 (1)} & \textbf{0.33 (1)} \\ 
   \hline
\end{tabular}
\vspace{-12pt}
\end{table}

Figure~\ref{fig:rq2torq5}(b) shows the best approaches ranked using \RELB~ by
their mean \RANKSCORE~ over both data sets. The mean value of \RELB~ and the
\RANKSCORE~ for each of these best approaches are listed in Table~\ref{tbl:results-relb}. 
There are actually ten approaches with a perfect \RANKSCORE~ of 1, i.e.,
Zimmermann09-SVM, Watanabe09-SVM, Turhan09-SVM, Ryu15-SVM, PHe15-SVM,
Peters15-SVM, Ma12-SVM, CamargoCruz09-SVM, Amasaki15-SVM, and the trivial
baseline FIX. Notably, all of the CPDP approaches use the SVM as classifier.
Herbold\etal~\cite{Herbold2017b} found in their benchmark, that unless the bias
towards non-defective instances is treatet, SVMs often yield trivial or nearly
trivial classifiers. We checked the raw results and found that this applies
here, too. All of these SVMs are either trivial classifiers predicting
only one class, or nearly trivial. Thus, they are nearly the same as the trivial baseline
FIX. Since we use the size of entities as tie-breaker, this means that ranking
code by size starting with the smallest instances until 20\% of
code is covered, is more effecient than actual \ac{CPDP}. When we investigate at
the actual values of \RELB, we can conclude that we find on average 32\%--35\%
percent of the defects that way, depending on the data set and which of the top ranked approaches is used. 

\begin{framed}
\noindent\textbf{Answer RQ3:} Trivial or nearly trivial predictions
perform best if only a small portion of the code undergoes additional quality
assurance. 
\end{framed}

\subsection*{RQ4: Which approach performs best independent of the prediction
threshold?}

\begin{table}[t]
\caption{Mean results for \AUCEC~ with \RANKSCORE~ in parenthesis.}
\label{tbl:results-aucec}
\vspace{-8pt}
\centering
\begin{tabular}{rll}\hline
\textbf{Approach} & \textbf{JURECZKO} & \textbf{AEEEM} \\ 
  \hline
ALL-SVM & \textbf{0.63 (1)} & 0.61 (0.95) \\ 
  Amasaki15-SVM & \textbf{0.63 (1)} & \textbf{0.63 (1)} \\ 
  CamargoCruz09-SVM & \textbf{0.63 (1)} & \textbf{0.63 (1)} \\ 
  Canfora13-MODEP100 & \textbf{0.62 (1)} & 0.61 (0.95) \\ 
  FZhang16-SC & 0.55 (0.6) & 0.57 (0.74) \\ 
  Herbold13-LR & 0.54 (0.6) & \textbf{0.63 (1)} \\ 
  Herbold13-SVM & 0.54 (0.6) & \textbf{0.62 (1)} \\ 
  Hosseini16-SBS & 0.48 (0.2) & 0.54 (0.63) \\ 
  Kawata15-SVM & \textbf{0.63 (1)} & 0.61 (0.95) \\ 
  Koshgoftaar08-SVM & \textbf{0.63 (1)} & \textbf{0.62 (1)} \\ 
  Liu10-GP & 0.57 (0.8) & 0.6 (0.79) \\ 
  Ma12-SVM & \textbf{0.63 (1)} & \textbf{0.61 (1)} \\ 
  Menzies11-SVM & \textbf{0.62 (1)} & \textbf{0.62 (1)} \\ 
  Nam15-SVM & 0.55 (0.6) & \textbf{0.62 (1)} \\ 
  Panichella14-CODEP-LR & \textbf{0.6 (1)} & 0.6 (0.95) \\ 
  Peters12-SVM & \textbf{0.63 (1)} & 0.6 (0.95) \\ 
  Peters13-SVM & \textbf{0.63 (1)} & 0.6 (0.95) \\ 
  Peters15-SVM & \textbf{0.62 (1)} & \textbf{0.62 (1)} \\ 
  PHe15-SVM & \textbf{0.63 (1)} & \textbf{0.61 (1)} \\ 
  RANDOM & 0.57 (0.6) & 0.55 (0.74) \\ 
  Ryu14-VCBSVM & 0.54 (0.6) & 0.6 (0.89) \\ 
  Ryu15-SVM & \textbf{0.62 (1)} & \textbf{0.61 (1)} \\ 
  Tantitham.16-SVMCARET & 0.55 (0.6) & \textbf{0.62 (1)} \\ 
  FIX & \textbf{0.63 (1)} & \textbf{0.61 (1)} \\ 
  Turhan09-SVM & \textbf{0.63 (1)} & 0.61 (0.95) \\ 
  Uchigaki12-LE & 0.46 (0.2) & 0.49 (0.11) \\ 
  Watanabe08-SVM & \textbf{0.62 (1)} & \textbf{0.62 (1)} \\ 
  YZhang15-MAXVOTE & 0.53 (0.6) & 0.54 (0.42) \\ 
  ZHe13-LR & 0.55 (0.6) & \textbf{0.61 (1)} \\ 
  ZHe13-NB & 0.52 (0.6) & \textbf{0.6 (1)} \\ 
  ZHe13-SVM & 0.54 (0.6) & \textbf{0.62 (1)} \\ 
  Zimmermann09-SVM & \textbf{0.62 (1)} & \textbf{0.62 (1)} \\ 
   \hline
\end{tabular}
\vspace{-12pt}
\end{table}

Figure~\ref{fig:rq2torq5}(c) shows the best approaches ranked using \AUCEC~ by their
mean \RANKSCORE~ over both data sets. The mean value of \AUCEC~ and the
\RANKSCORE~ for each of these best approaches are listed in
Table~\ref{tbl:results-aucec}. The results are very similar to the results from
RQ3, i.e., we have a large group of approaches with a perfect \RANKSCORE~ of 1,
all of which use SVM as classifiers including the baseline FIX. Thus, starting
with small code entities is again the best strategy. 

\begin{framed}
\noindent\textbf{Answer RQ4:} Trivial or nearly trivial predictions perform best
without a fix prediction threshold.
\end{framed}

\subsection*{RQ5: Which approach performs best overall?}

Figure~\ref{fig:rq2torq5}(d) shows the best approches ranked using the three
metrics \NECM, \RELB, and \AUCEC~ over both data sets. Given the results for RQ2-RQ4, it
is not suprising that the best ranking approach is the trivial baseline FIX with
a nearly perfect mean \RANKSCORE~ of 0.994. The best \ac{CPDP} approaches are
Canfora13-MODEP100 with a mean \RANKSCORE~ of 0.969 and Liu10-GP with a mean
\RANKSCORE~ of 0.888. 

\begin{framed}
\noindent\textbf{Answer RQ5:} No \ac{CPDP} approach outperforms our trivial
baseline on average over three performance metrics. The best performing
\ac{CPDP} approaches are Canfora13-MODEP100 and Liu10-GP.
\end{framed}

\subsection*{RQ6: Is the overall ranking based on cost metrics
different from the overall ranking based on the AUC, F-measure, G-Measure, and
MCC?}

Figure~\ref{fig:rq6} shows the \RANKSCORES~ of the best ranked approaches from
RQ5, but ranked with \AUC, \FMEAS, \GMEAS, and \MCC~ instead, i.e., the
metrics used by Herbold\etal~\cite{Herbold2017b, Herbold2017d} for ranking with machine
learning metrics that do not consider costs. In case the rankings are correlated, we
would expect that the \RANKSCORES~ are roughly sorted in descending order from
top to bottom. However, this is clearly not the case. The two top ranking
approaches with cost metrics, i.e., FIX and Canfora13-MODEP have both very low
\RANKSCORES~ with the metrics used by Herbold\etal, the next ranking approaches
are much better with at least mediocore \RANKSCORES. Please note that the
\RANKSCORE~ values here are not the same as determined by
Herbold\etal~\cite{Herbold2017b, Herbold2017d}, because we only use two of the
five data sets for this comparison. We confirmed this visual observation using
Kendall's $\tau$ as correlation measure  We observe almost no correlation of
$\tau=-0.047$ between both rankings.

\begin{framed}
\noindent\textbf{Answer RQ6:} The cost-sensitive ranking is completely
different from the ranking base on \AUC, \FMEAS, \GMEAS, and \MCC. Thus,
machine learning metrics are unsuited to predict the cost
efficiency of approaches, and cost metrics are likewise unsuited to
predict the performance measured with machine learning metrics. 
\end{framed}

\section{Discussion}
\label{sec:discussion}

Our results cast a relatively devastating light on the cost efficiency
of \ac{CPDP}. Based on our results, it just seems
better to follow a trivial approach and assume everything as equally likely to
contain defects. What we find notable is that
most of the state of the art of \ac{CPDP} has ignored cost
metrics. Only Khoshgoftaar08, Liu10, Uchigaki12, Canfora13, Panichella14, and YZhang15,
i.e, six out of 26 approaches used any effort or cost related metrics when
evaluating their work. Only two of these approaches, i.e., Liu10 and Canfora13
optimize for costs. These are also the
two best ranked approaches after the trivial baseline if we use all three
cost metrics. Liu10 weights false negatives fifteen
times stronger than false positives, i.e., puts a strong incentive on
identifying defective instances in comparison to misidentifying instances as defective. This
incentive means that they optimize the metric \NECM. As a results, Liu10-GP is better ranked than the trivial baseline,
which is the only time this happend for all approaches and metrics in our
benchmark. Canfora13 use two objectives for optimization: the
\RECALL\footnote{Percentage of predicted defective instances.} and the effort in
lines of code considered. This is similar to \AUCEC, but not sufficient to
outperform our trivial baseline even for that metric. Thus, optimizing for
cost directly seems to be vital for \ac{CPDP} if the created models should be cost efficient.

Another interesting aspect of our findings is the lack of correlation between
the ranking using cost metrics and machine learning metrics. One
would expect that good classification models in terms of machine learning
metrics, also perform well under cost considerations. 
We believe that this
correlation is missing because the performance of the \ac{CPDP} models
is too bad, especially the \PRECISION,\footnote{Percentage of defective
predictions that are actually not defective.} but also in terms of
\RECALL.
According to the benchmark from Herbold\etal~\cite{Herbold2017d}, only very few
predictions achieve a $\RECALL\geq0.7$ and $\PRECISION\geq0.5$ at the same time. 
In other words, finding 70\% of the defects with at least 50\% of the
predicted instances being actually defective is almost impossible with the state
of the art of \ac{CPDP}. Herbold\etal~actually found that the baseline FIX is
one of the best approaches when it comes to these criteria. Thus, this trivial
baseline, which ranks badly in the benchmark by Herbold\etal, performs well in
our cost setting, because it has a very high recall.
We think that this explains the lack of correlation. If the general performance of \ac{CPDP} models would be higher, we
believe that it is likely that the results would be correlated to costs.

Both the machine learning metric benchmark by Herbold\etal~\cite{Herbold2017b}
and our cost-sensitive benchmark share the trait
that relatively na\"{i}ve baselines outperform many approaches from the state
of the art and that the best ranking \ac{CPDP} approach is relatively old. In
Herbold\etal's\cite{Herbold2017b} benchmark, the baseline is ALL, i.e., using
all data without any treatment for classification. In our case it is the trivial
baseline FIX. Moreover, in both benchmarks approaches that are over 5 years old
often perform better than most newer approaches. While most researchers
implement comparisons to baselines, other approaches from the state of the art
are only seldomly replicated. Furthermore, the amount of data is important for
the generalization of results. For example,
Canfora\etal~\cite{Canfora2015} actually compare their results with the same
trivial baseline we do. The difference between their work and ours is that we
use more data for our analysis. They use a subset of 10
products from the JURECZKO data, whereas we use 62 products. Together, this
shows that diligent comparisons to the state of the art and baselines are vital,
but also that using large data sets is mandatory to reach firm conclusions.
Otherwise, many approaches will be published, which only marginally advance the
state of the art, or only under very specific conditions, but not in a broader
context. Please note that while we use 67 products in total, this is still only
a small amount of data, in comparison to the thousands of Java projects on
GitHub alone that provide sufficient data in issue tracking system and commit
comments to be used for defect prediction studies. Therefore, we believe that
this sample size is major threat to the validity of this benchmark and defect
prediction research in general. The problem of small sample sizes was already
identified as a threat to the validity of repository mining studies in
general~\cite{Trautsch2016, Trautsch2017}. Without scaling up our sample sizes,
we may run into a serious replication crisis.

\section{Threats to validity}
\label{sec:threats}

Our benchmark has several threats to its validity. We distinguish between
internal validity, construct validity, and external validity. 

\subsection{Internal Validity}
Because we replicated only existing work, we do not see any internal threats to
the validity of our results. 

\subsection{Construct Validity}
The benchmark's construction influences the results. Threats to the validity of
the construction include unsuitable choice of performance metrics for the
research questions, unsuitable statistical tests, noisy or mislabeled
data~\cite{Jureczko2010}, as well as defects in our implementations.
To address these threats, we based our metrics and statistical tests on the
literature, used multiple data sets to mitigate the impact of noise and
mislabeling, and tested all implementations. Moreover, we performed a sanity
check against the results from Herbold\etal~\cite{Herbold2017}, assuming that
the construct of that benchmark is valid.

\subsection{External Validity}
The biggest threat to the external validity is the sample size of software
products. Within this benchmark, we consider 67 products, which is large for
defect prediction studies, but small in relation to the overall number of
software projects. Therefore, we cannot firmly conclude that we found general
effects and not random effects that only hold on our used data. 
\section{Conclusion}
\label{sec:conclusion}

Within this paper, we present a benchmark on \ac{CPDP} using effort and cost
metrics. We replicated 26 approaches from the state of the art published between
2008 and 2016. Our results show that 
a trivial approach that predicts everything as defective performs better than
the state of the art for \ac{CPDP} under cost considerations.
The two best \ac{CPDP}  were proposed by Liu\etal~\cite{Liu2010} and
Canfora\etal~\cite{Canfora2013, Canfora2015} and are close to the
trivial predictions in performance. These are also the only two approaches, that
directly optimize costs. We
suspect that the generally insufficient performance of \ac{CPDP} models, that
was already determined in another benchmark by
Herbold\etal~\cite{Herbold2017b}, is the reason for the bad performance of
\ac{CPDP} in a cost-sensitive setting. It seems that optimizing directly for
cost and not performance of the prediction model is currently the only way to
produce relatively cost-efficient \ac{CPDP} models.

In our future work, we will build on the findings of this benchmark and
use the gained insights to advance the state of the art. We plan to define a
general \ac{CPDP} framework, that will allow a better selection of optimization criteria for
approaches. We want to see if it is possible to build a wrapper
around the defect prediction models that can make them optimize for costs, e.g.,
by injecting other performance metrics into the machine learning algorithms as
optimization criteria, or manipulate the training data such that
performance estimations, e.g., based on the \ERROR~ are more similar to actual
costs. We hope to advance the state of the art this way to be more
cost-efficient, such that \ac{CPDP} becomes a significant improvement in
comparison to assuming everything is defective. In parallel to this, we will
collect more defect prediction data in order to scale up the sample size and
allow better conclusions about the generalizability of defect prediction results and reduce this major threat to the validity of
defect prediction research.

\begin{acks}
The authors would like to thank GWDG for the access to the scientific compute
cluster used for the training and evaluation of thousands of defect prediction
models.
\end{acks}

\bibliographystyle{ACM-Reference-Format}
\bibliography{literature} 

\end{document}